\documentclass[10pt,conference]{IEEEtran}
\usepackage{cite}
\usepackage{amsmath,amssymb,amsfonts}
\usepackage{algorithmic}
\usepackage{graphicx}
\usepackage{textcomp}
\usepackage{xcolor}
\usepackage[hyphens]{url}

\usepackage{algorithm}
\usepackage{caption}
\usepackage{subcaption}
\usepackage{authblk}
\usepackage[english]{babel}
\usepackage{amsthm}
\usepackage{bm}
\usepackage{svg}
\usepackage{float}
\usepackage{multirow}

\def\BibTeX{{\rm B\kern-.05em{\sc i\kern-.025em b}\kern-.08em
    T\kern-.1667em\lower.7ex\hbox{E}\kern-.125emX}}

\title{
    Making Neural Networks More Suitable \\for Approximate Clifford+$T$ Circuit Synthesis
}
\author[1]{Mathias Weiden}
\author[1]{Justin Kalloor}
\author[1]{John Kubiatowicz}
\author[2]{Costin Iancu}
\affil[1]{Department of Electrical Engineering and Computer Science, University of California, Berkeley}
\affil[ ]{\textit{\{mtweiden, jkalloor3, kubitron\}@cs.berkeley.edu}}
\affil[2]{Computational Research Division, Lawrence Berkeley National Laboratory}
\affil[ ]{\textit{cciancu@lbl.gov}}

\newtheorem{definition}{Definition}[section]

\begin{document}
\newcommand{\unitarygroup}[0]{\mathbb{U}(2^n)}
\newcommand{\realnum}[0]{\mathbb{R}}
\newcommand{\complexnum}[0]{\mathbb{C}}
\newcommand{\phase}[1]{e^{i#1}}
\newcommand{\expectation}[0]{\mathbb{E}}
\newcommand{\integers}[0]{\mathbb{Z}}

\newcommand{\target}[0]{U_{tar}}
\newcommand{\circuit}[0]{C_t}
\newcommand{\perturbation}[0]{U_{\epsilon}}

\newcommand{\initialstates}[0]{S_{I}}
\newcommand{\terminalstates}[0]{S_{T}}
\newcommand{\statespace}[0]{S}
\newcommand{\actionspace}[0]{A}
\newcommand{\s}[0]{s}
\newcommand{\mdpstate}[1]{\s_{#1}}
\newcommand{\ac}[0]{a}
\newcommand{\action}[1]{\ac_{#1}}
\newcommand{\trajectory}[0]{\tau}

\newcommand{\gridsynth}[0]{\text{\emph{gridsynth}}}

\long\def\comment#1{}
\long\def\note#1{{\em #1 }}
\def\parah#1{\vspace*{0.0in} \noindent{\bf #1:}}
\newcommand{\red}[1]{\textcolor{red}{#1}}

\maketitle
\thispagestyle{plain}
\pagestyle{plain}


\begin{abstract}
Machine Learning with deep neural networks has transformed computational approaches to scientific and engineering problems.
Central to many of these advancements are precisely tuned neural architectures that are tailored to the domains in which they are used.
In this work, we develop deep learning techniques and architectural modifications that improve performance on reinforcement learning guided quantum circuit synthesis—the task of constructing a circuit that implements a given unitary matrix.
First, we propose a global phase invariance operation which makes our architecture resilient to complex global phase shifts.
Second, we demonstrate how augmenting data with small random unitary perturbations during training enables more robust learning.
Finally, we show how encoding numerical data with techniques from image processing allow networks to better detect small but significant changes in data.
Our work enables deep learning approaches to better synthesize quantum circuits that implement unitary matrices.
\end{abstract}

\section{Introduction}
\label{sec:intro}
Effective compilation and optimization of quantum programs is crucial for both near-term and fault-tolerant quantum computing.
Recent advances in fields such as program synthesis~\cite{chen_2020_programsynthesis, shen2022benchmarkinglanguagemodelscode} have demonstrated the power of machine learning in classical compilation flows, driving interest in its application to compiling quantum programs~\cite{alexeev_2024_ai_in_qc}.
Many compilation techniques in this domain focus on unitary representations of quantum circuits and operations.
While several works have shown how machine learning can be applied to these problems, little attention has been given to tailoring machine learning techniques specifically for quantum compilation.

We introduce three architectural modifications to neural networks which make them more suitable for building quantum circuits which approximately implement a given function.
The postulates of quantum mechanics imply that differences in global complex phase do not affect the underlying function of quantum operators~\cite{mike_and_ike}.
Incorporating this property at the architectural level improves robustness against different numerical representations of the same unitary.
Additionally, circuit synthesis often demands some level of approximation.
To accommodate this, we introduce a unitary-perturbative data augmentation technique which enables neural networks to associate learned quantities with closely related unitaries rather than individual matrices.
Lastly, we adopt data encoding methods from 3D image processing which makes neural networks better at detecting small changes in numerical data.

The remainder of the paper is organized as follows:
In Section~\ref{sec:background} we provide the necessary background and review related works.
Section~\ref{sec:method} details the architectural modifications.
In Section~\ref{sec:experiments} we present experimental evaluations of these modifications on the task of quantum circuit synthesis.
We discuss the implications of our findings in Section~\ref{sec:discussion} and conclude with final remarks in Section~\ref{sec:conclusion}.

\section{Background}
\label{sec:background}
In this section, we present background on the role unitary matrices play in quantum computing.
We describe the problem of \emph{quantum circuit synthesis} with the universal and Fault-Tolerant (FT) Clifford+$T$ gate set.
We finish this section by discussing related work which use machine learning to address these problems, and work that targets learning on inputs which consist of complex numbers.

\subsection{The role of unitaries in quantum computing}
Quantum programs transform input quantum state vectors to output quantum state vectors.
The input to a quantum program is typically assumed to be the zero state vector:
\[
    |0\rangle = \begin{bmatrix} 1 & 0 & \dots & 0\end{bmatrix}^T \in \complexnum^{2^n}
\]
where, $n$ is the number of qubits in the system.
The dynamics of the quantum program can be described by a unitary matrix
\[
    U \in \unitarygroup \text{  such that  } U|0\rangle = |\psi\rangle
\]
for some desired output state vector $|\psi\rangle \in \complexnum^{2^n}$.

The similarity between two unitaries, $A$ and $B$ can be quantified using the Hilbert-Schmidt distance
\begin{equation}
    \label{eqn:distance}
    d_{HS}(A, B) = \sqrt{1 - \frac{1}{4^n} |Tr(A B^\dagger)|^2 }
\end{equation}
Notice how if two unitaries differ only by a global phase, meaning $B = e^{i\theta}A$ for $\theta \in \realnum$, then $d_{HS}(A, e^{i\theta}A) = 0$.
This is a desirable trait for a distance function to have because unitaries that differ by global phases are indistinguishable using projective measurements~\cite{mike_and_ike}.

The trick of quantum compilation is to find some quantum circuit, or a sequence of quantum gates, which implements the desired unitary.
In many cases, this unitary matrix cannot be implemented exactly and must be approximated~\cite{solovay_kitaev}.
The quantum gates which make up quantum circuits can be categorized in several ways.
These include single-qubit gates, such as
\begin{equation}
    \label{eqn:one_qubit_clifford}
    H = \frac{1}{\sqrt{2}} \begin{bmatrix} 1 & 1 \\ 1 & -1 \end{bmatrix}, \quad S = \begin{bmatrix} 1 & 0 \\ 0 & i \end{bmatrix}
\end{equation}
and two-qubit entangling gates, like
\begin{equation}
    \label{eqn:two_qubit_clifford}
    CNOT = \begin{bmatrix} 1 & 0 & 0 & 0 \\ 0 & 1 & 0 & 0 \\ 0 & 0 & 0 & 1 \\ 0 & 0 & 1 & 0 \end{bmatrix}, \quad
    CZ = \begin{bmatrix} 1 & 0 & 0 & 0 \\ 0 & 1 & 0 & 0 \\ 0 & 0 & 1 & 0 \\ 0 & 0 & 0 & -1 \end{bmatrix}
\end{equation}
The gates presented in Equations~\ref{eqn:one_qubit_clifford} and~\ref{eqn:two_qubit_clifford} are members of the Clifford group, and not sufficient for universal quantum computation~\cite{mike_and_ike}.
Universal quantum computation requires additional non-Clifford gates such as
\begin{equation}
    \label{eqn:one_qubit_nonclifford}
    T = \begin{bmatrix} 1 & 0 \\ 0 & e^{i\pi/4} \end{bmatrix} \quad
    R_Z(\phi) = \begin{bmatrix} 1 & 0 \\ 0 & e^{i\phi} \end{bmatrix}
\end{equation}
With the $R_Z$ gate, we see how non-Clifford gates may be \emph{parameterized} by real numbers.

We call the set of gates $\{H, S, CZ, T\}$ the Clifford+$T$ gate set.
This gate set is both universal~\cite{solovay_kitaev} and fault-tolerant~\cite{mike_and_ike}.
As an extension to the Clifford+$T$ gate set, we can consider the Clifford+$\phi$ gate set, which consists of $\{H, S, CZ, T, R_Z(\phi)\}$ gates.
Optimal methods exist for decomposing single-qubit $R_Z$ gates into sequences of Clifford+$T$ gates~\cite{ross_2016_gridsynth, bocharov_2015_rus, bocharov_2015_fallback}, so any circuit expressed in Clifford+$\phi$ can also be expressed in Clifford+$T$ gates.

\subsection{Quantum Circuit Synthesis}
Sometimes we are given a unitary $U$ that corresponds to some useful quantum program, but we do not have a sequence of quantum gates which implements $U$.
We can use a \emph{unitary synthesis algorithm} to construct such a sequence of gates.
More concretely, we want to find a circuit $\circuit$ which satisfies the equation
\[
    d_{HS}(\circuit, U) \leq \epsilon
\]
for our target unitary $U$.
The hyperparameter $\epsilon \ll 1$ determines how closely $\circuit$ approximates $U$.

Every $n$-qubit circuit $\circuit$ can be described as a sequence of primitive gates taken from a finite set $\actionspace \subset \unitarygroup$.
Every primitive gate has a known associated unitary $\ac \in \actionspace$.
Each gate's unitary representation $\ac$ depends both on the gate's function and the qubit it acts on.
A circuit consisting of $t$ gates is associated with a unitary matrix
\begin{equation}
    \label{eqn:circuitunitary}
    \circuit = \action{t} \times \dots \times \action{1}.
\end{equation}

\begin{figure}[t]
    \centering
    \includegraphics[width=0.48\textwidth]{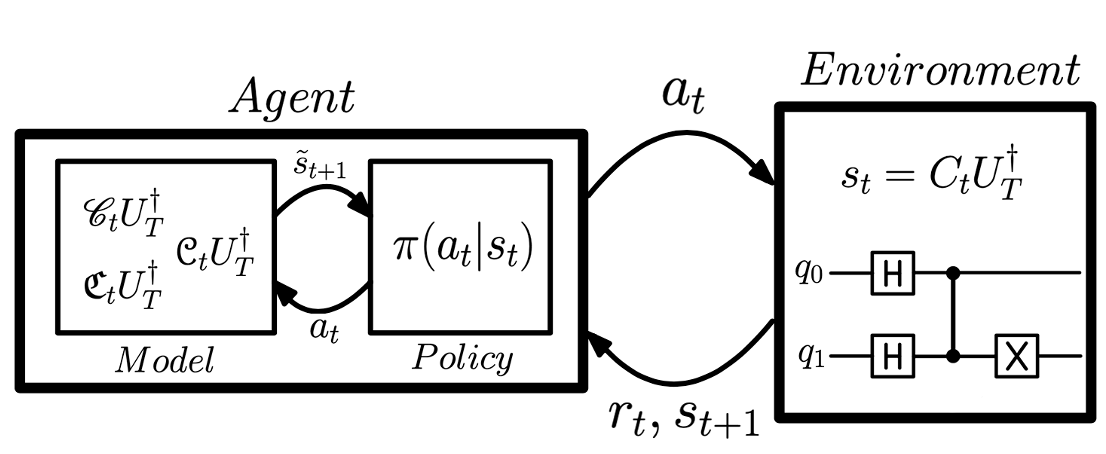}
    \caption{
        A model-based Reinforcement Learning agent. The agent observes the state of the environment $\mdpstate{t}$, uses the policy to commit to an action $\action{t}$, then receives a reward $r_t$.
        Model-based synthesis approaches can evaluate the value of many actions before committing to them.
    }
    \label{fig:mbrl}
\end{figure}
\subsection{Quantum Circuit Synthesis with Reinforcement Learning}
At a high-level, a Reinforcement Learning (RL) algorithm uses interactions with an environment to train an agent (some entity which makes decisions) to select actions which maximizes the long term reward that the agent receives.
This is illustrated in Figure~\ref{fig:mbrl}.

The process of constructing a quantum circuit which implements a unitary matrix can be described using a \emph{Markov Decision Process} (MDP)~\cite{zhang_2020_topological_compiling, moro_2021_drlcompiling, chen_2022_efficient, alam_2023_synthesis_mdp, rietsch_2024_unitary}.
We define the MDP as a tuple $(\statespace, \initialstates, \terminalstates, \actionspace, r, p)$.
Here $\statespace$ is a set of states, $\initialstates$ a set of initial states, $\terminalstates$ a set of terminal states, $\actionspace$ a set of actions or gates, $r: \statespace \times \actionspace \rightarrow \realnum$ a reward function, and $p$ a distribution which assigns a probability to next states given a state-action pair.
We use the term state (or sometimes unitary state) to describe an RL state $s_t$, not a quantum mechanical state vector or density operator.

Given an MDP, an RL algorithm optimizes for a policy which maximizes the total expected cumulative rewards.
Formally, a policy denoted as $\pi(\action{t} | \mdpstate{t})$ assigns a probability to each possible action $\action{t} \in \actionspace$ given state $\mdpstate{t}$. A state $\mdpstate{t}$ has value
\begin{equation}
    V^\pi(\mdpstate{t}) = \expectation_\pi \sum_{k=0}^{T} \gamma^{k} r(\mdpstate{t + k}, \action{t + k})
    \label{eqn:valuefunc}
\end{equation} 
where $\gamma \in (0,1]$ is a discount factor. The value is the expected cumulative discounted reward if policy $\pi$ is followed. We can additionally define the state action value or \emph{Q function}
\begin{equation}
    Q^\pi (\mdpstate{t}, \action{t}) = r(\mdpstate{t}, \action{t}) + \gamma \expectation_{p} V^\pi(\mdpstate{t+1})
    \label{eqn:qfunc}
\end{equation}
which is the value of taking action $\action{t}$ when in state $\mdpstate{t}$ and then following policy $\pi$.
Value functions differ from Q functions in that they only require a state, not a state-action pair.
More details relating to value functions can be found in~\cite{suttonbarto}.

The agent is trained to maximize one of these value functions, which directly corresponds to maximizing the long term expected reward that the agent will receive from the environment.
We use the binary reward function
\begin{equation}
    \label{eqn:reward}
    r(\mdpstate{t}) = \begin{cases} 
      +1 & \text{if } \mdpstate{t} \in \terminalstates \\
      -1 & \text{else}
   \end{cases}
\end{equation}
to train the agent to find low gate count implementations of unitaries.

\subsection{Related Work}
The vast majority of machine learning models rely on real number arithmetic for computation and consider only real number valued inputs.
In some domains, like satellite imaging~\cite{singhal_2021_codomain, singhal_2022_spectral} and Magnetic Resonance Imaging (MRI)~\cite{virtue_2019_complex}, complex valued input data is ubiquitous.
This works takes inspiration primarily from these works, as unitary quantum circuit synthesis requires taking complex valued unitary matrices as input.

\subsubsection{Machine Learning in Quantum Compilation}
Much work has been done on how Machine Learning may be used to help compile quantum programs and manage quantum computers.
A recent survey has touched on several ways ML may be leveraged in this field~\cite{alexeev_2024_ai_in_qc}.
Ideas span from quantum error decoding~\cite{nachmani_learning_2016, krastanov_deep_2017, bausch_learning_2024}, circuit optimization~\cite{quetschlich_2022_compiler_optimization, li_2024_quarl}, to quantum circuit synthesis and beyond.

One work has shown how machine learning can accelerate existing synthesis workflows~\cite{weiden_2023_qseed}.
Most work however has focused on how Reinforcement Learning can be applied to this problem~\cite{zhang_2020_topological_compiling, moro_2021_drlcompiling, chen_2022_efficient, rietsch_2024_unitary}.
This work does not compete directly with the ideas presented in these later works, but instead proposes modifications to the neural architecture used which can improve synthesis performance.
The relative performance gains when using our techniques along with the techniques presented in those works can be inferred from the data in Section~\ref{sec:experiments}.

\subsubsection{Synthesis by Diagonalization}
A recent paper has demonstrated that the class of unitaries which can be effectively synthesized by search-based synthesis algorithms (like Reinforcement Learning) can be expanded by loosening the synthesis objective from matrix inversion to diagonalization~\cite{weiden_2025_diagonalization}.
We adopt this paradigm for several experiments presented in Section~\ref{sec:experiments}.
Doing so requires using the ``diagonal distance'' detailed in Section 4.1 of that paper.
More details on how this relates to synthesis with Reinforcement Learning are discussed in Section~\ref{sec:experiments}.


\section{A Neural Architecture for Circuit Synthesis}
\label{sec:method}




We present an architecture which we have called the Deep Value Network for Unitaries (DVNU).
DVNU indirectly learns a synthesis policy $\pi$ using Value Iteration \cite{bellman_1957_dynamicprogramming}.
Compared to the approaches described in \cite{moro_2021_drlcompiling, chen_2022_efficient, rietsch_2024_unitary}, our model-based method trains faster and achieves better synthesis performance. 

We introduce advancements necessary for compiling more complex quantum programs.
First, we introduce a global phase invariance operation.
Invariance to global phase ensures that networks learn to associate values with a canonical form of a unitary state.
Second, we introduce a unitary perturbation data augmentation.
By applying small unitary disturbances to states during training, the network learns to associate values with regions rather than points on the Bloch sphere. 
Finally, we adopt a data encoding technique from the field of 3D image processing to better capture small changes in numerical data.
The impact of these architectural changes are illustrated in Figure~\ref{fig:invar_depiction}.
An ablation study (i.e. an experiment where we selectively remove each modification and train) summarizes the effect on performance of all architectural changes (see Figure~\ref{fig:ablation}).

\subsection{Modeling Synthesis}
Our search-based synthesis method relies on a form of RL called \emph{model}-based learning.
\begin{definition}
    A \emph{model} is a function $f$ which predicts the dynamics of a controllable environment $f(\mdpstate{t}, \action{t}) = \tilde{\s}_{t+1}$.
\end{definition}
Model-based RL often assumes that a neural network learns system dynamics.
In the case of quantum circuit synthesis, knowledge of the state $\mdpstate{t}$ and the action $\action{t}$ allows for exact, deterministic, knowledge of the system dynamics because $\mdpstate{t+1} = \action{t} \times \mdpstate{t}$ or $\mdpstate{t+1} = \mdpstate{t} \times \action{t}$, so no learning is needed.

\subsection{Global Phase Invariance}
\label{sec:invariantarch}
\begin{figure}[t]
    \centering
    \includegraphics[width=0.5\textwidth]{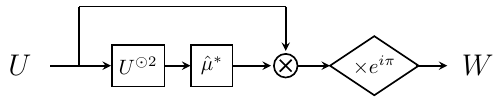}
    \caption{
        A unitary $U$ can be transformed to global phase invariant representation $W$ by $1)$ squaring $U$ element-wise, $2)$ computing and conjugating the normalized mean to obtain a unit magnitude complex number $\hat{\mu}^*$, $3)$ multiplying $\hat{\mu}^* \times U$, and $4)$ conditionally correcting the phase by an angle of $\pi$. The Appendix contains further details.
    }
    \label{fig:invariantarch}
\end{figure}

Two observables which differ only by a global phase cannot be distinguished experimentally \cite{mike_and_ike}.
When dealing with unitary operators numerically, we are forced to choose representations which may differ due to global phase.
To ensure that equivalent unitaries are represented in the same way, we require global phase invariance.
\begin{definition}
    A function $g$ is said to be \emph{global phase invariant} if for two unitaries $U, W \in \unitarygroup$ and $\theta \in \realnum$ such that $U = \phase{\theta}W$, it is the case that $g(U) = g(W)$.
\end{definition}
We construct a new phase invariant neural network layer similar to the manner of \cite{singhal_2021_codomain}, which showed how nonlinear activation functions may be made equivariant to complex phase.
Figure~\ref{fig:invariantarch} illustrates this process.

\begin{figure*}[th]
    \centering
    \includegraphics[width=0.7\textwidth]{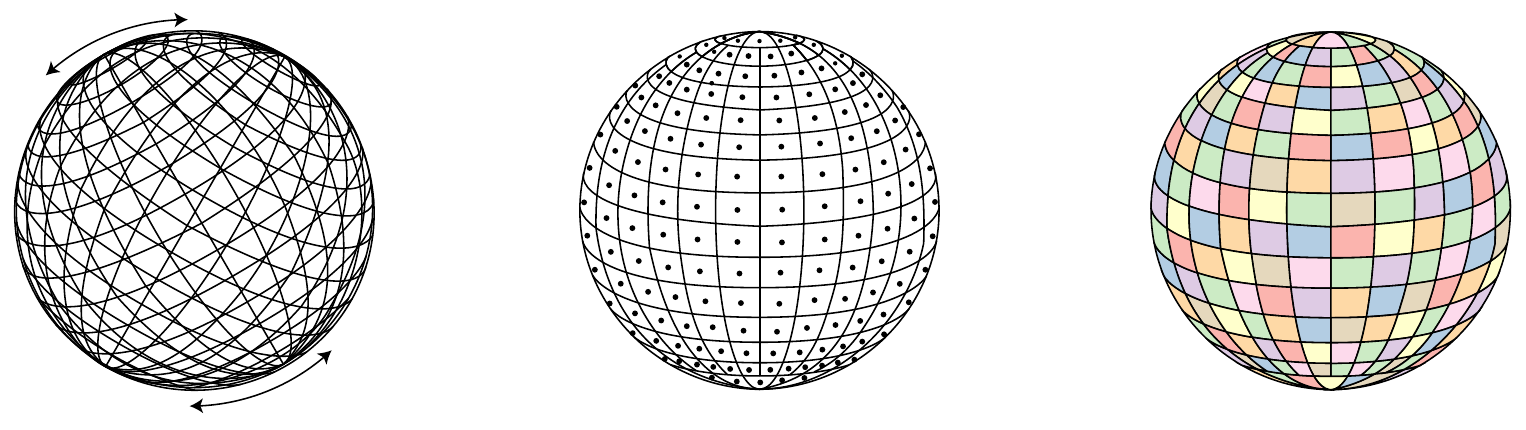}
    \caption{
        Visualization of global phase invariance and unitary perturbative data augmentations.
        Points on the \emph{Bloch sphere} represent pure single-qubit quantum mechanical state vectors \cite{mike_and_ike}.
        \emph{Left:} Without global phase invariance, the Bloch sphere is rotated so that equivalent quantum states have different representations.
        \emph{Center:} With no unitary perturbations on the inputs, reachable quantum states are only represented as single points.
        \emph{Right:} Global phase invariance and unitary perturbations ensure that quantum states are represented consistently and that the network learns to associate entire $\epsilon$-regions with values.
    }
    \label{fig:invar_depiction}
\end{figure*}

For unitary $U$, we define $\mu = \frac{1}{N^2}\sum u_{ij} \in \complexnum$ to be the element-wise mean.
As it may be the case that $\mu=0$, we consider the mean of the element-wise squared matrix $U^{\odot 2}$ which is $\mu^\prime = \frac{1}{N^2}\sum u_{ij}^2$.
We normalize this to obtain $\hat{\mu^\prime} = \mu^\prime / \|\mu^\prime\|$.
Next we halve the phase of $\hat{\mu}^\prime$ and conjugate to receive $\hat{\mu}^*$.
Finally we multiply by $U$ and conditionally add a phase correction of $\pi$ to receive a global phase invariant representation $W$.
This process is depicted in Figure~\ref{fig:invariantarch}.
\begin{proof}
    Say $U = \phase{\theta}W$. The element-wise mean of $U^{\odot2}$ is $\mu^\prime = \phase{(2\theta + 2\pi k)}x$. Normalizing, halving the phase, and conjugating gives $\hat{\mu}^* = e^{-i(\theta + k\pi)}$. Multiplying $U$ by the conjugated normalized mean gives $\hat{\mu}^* U = \phase{\theta} e^{-i(\theta + k\pi)} W = e^{-ik\pi}W$. From this final result we cannot detect whether $k$ is odd or even. We choose to ensure that the first maximum magnitude entry of the first row of $W$ should have a positive real component. If it does not, we apply the $\pi$ phase correction.
\end{proof}

\subsection{Unitary Perturbations for Robust Value Learning}
\label{sec:perturbations}
Unitary synthesis constructs approximate implementations of unitaries.
As indicated by the distance condition in Equation~\ref{eqn:distance}, each unitary is associated with an $\epsilon$-disk of unitaries (all unitaries $V$ which satisfy $d_{HS}(U, V) \leq \epsilon$) which are considered approximately equivalent.
This fact naturally inspires a perturbative data augmentation technique: small unitary disturbances are applied to unitary states in order to learn this approximate equivalence structure.

Concretely, given a unitary state $\mdpstate{t}$, a unitary perturbation $\perturbation$ satisfying $d_{HS}(\perturbation, I) < \epsilon$ is applied so that the state viewed by the agent is $\perturbation \times \mdpstate{t}$ (or $\mdpstate{t} \times \perturbation$).
A visualization of the unitary perturbation data augmentation is illustrated in Figure~\ref{fig:invar_depiction}.
The impact on performance is depicted in Figure~\ref{fig:ablation}.

Applying a unitary perturbation $U_\epsilon$ to a unitary $V$ results in a unitary $W$ which is bounded by $d_{HS}(W, V) \leq \epsilon$.
\begin{proof}
    By Equation 1 of~\cite{wang_1994_trace_inequality}, we know that $\sqrt{1 - tr(U_\epsilon VV^\dagger)^2} \leq \sqrt{1-tr(U\epsilon)^2} + \sqrt{1-tr(VV^\dagger)}$. Because $d_{HS}(U_\epsilon, I) \leq \epsilon$, and $d_{HS}(V, V^\dagger) = 0$, we can conclude that $\sqrt{1 - tr(U_\epsilon VV^\dagger)^2} \leq \epsilon$. This implies that $d_{HS}(U_\epsilon V, V) = d_{HS}(W, V) \leq \epsilon$.
\end{proof}

\subsection{Neural Radiance Field Encoding}
\label{sec:nerf}
Unitary states are represented using $2 \times 2^n \times 2^n$ real tensors, where $n$ is the number of qubits. As we focus only on cases of high-precision synthesis where $n\in \{1, 2, 3\}$, our input data is extremely information dense. 
Past work has shown that neural networks struggle to fit high frequency patterns in data \cite{rahaman_2019_spectralbias, xu_2020_hifreq}.
We adopt a positional encoding technique used widely in the field of novel view synthesis and introduced by the NeRF project  \cite{mildenhall_2020_nerf}.
Input values are transformed by a map $\gamma: \realnum \rightarrow \realnum^{2L}$ such that
\begin{equation}
\begin{split}
    \gamma(x_i) = \Big(\sin(2^0 \pi x_i), \cos(2^0 \pi x_i), \dots, \\
    \sin(2^{L-1} \pi x_i), \cos(2^{L-1} \pi x_i) \Big)
\end{split}
\end{equation}
This encoding map plays a vital role in allowing networks to detect small magnitude differences in each element of a unitary state. The encoding therefore transforms input tensors from a shape of $(2, 2^n, 2^n)$ to a shape of $(2, 2^n, 2^n, 2L)$. For our experiments, we set $L=15$.

\subsection{Deep Value Network for Unitary Synthesis}
\label{sec:ampcapproach}

We leverage deep value iteration to learn gate set specific value functions. 
We refer to this architecture as the Deep Value Network for Unitaries (DVNU).
This contrasts with past work such as \cite{moro_2021_drlcompiling, chen_2022_efficient} which leverage Deep Q Networks (DQN) \cite{mnih_2013_dqn}.
To train a DQN agent, a state-action value (Equation~\ref{eqn:qfunc}) must be learned for \emph{each} $(s_t, a_t)$ pair.
To ensure adequate training, the agent must receive examples of many different actions taken in each state.
When the number of actions (in this case the gate set $|\actionspace|$) is large, the sample inefficiency of the DQN algorithm limits training progress.
In comparison, the DVNU agent must only learn the value of a state $s_t$.
The policy used is 
\begin{equation}
    \pi(\action{} | \mdpstate{t}) = \arg \max_{\action{} \in \actionspace} V^\pi_\theta(\action{} \times \mdpstate{t})
\end{equation}
Value iteration \cite{suttonbarto} is possible because synthesis can be modeled exactly and the reward is a deterministic function of the state.
The previous work of \cite{zhang_2020_topological_compiling} also employed a method based on value iteration for topological quantum computing.
More recently \cite{alam_2023_synthesis_mdp} demonstrated how in restricted cases, tabular (as opposed to deep RL) policy iteration can be used to solve for optimal synthesis policies.
This architecture is also well suited for scenarios where synthesis is framed as a \emph{Monte Carlo Tree Search} problem~\cite{rietsch_2024_unitary}, as this technique typically involves training a value network.

\begin{figure*}
    \centering
    \begin{subfigure}{\columnwidth}
        \includegraphics[width=\linewidth]{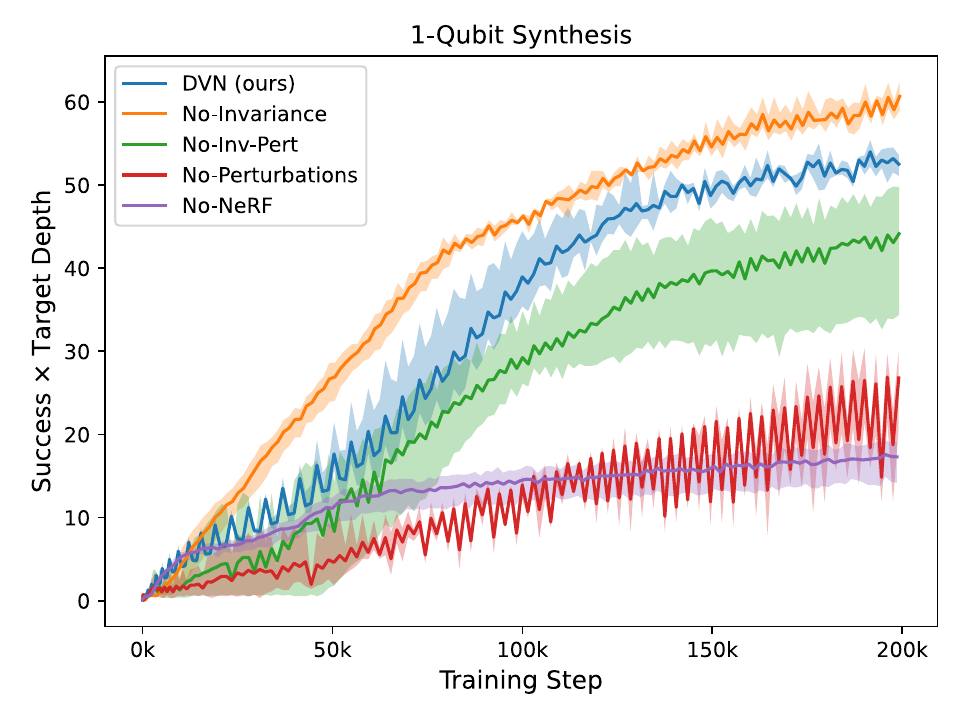}
        \label{fig:ablation_q1}
    \end{subfigure}
    \begin{subfigure}{\columnwidth}
        \includegraphics[width=\linewidth]{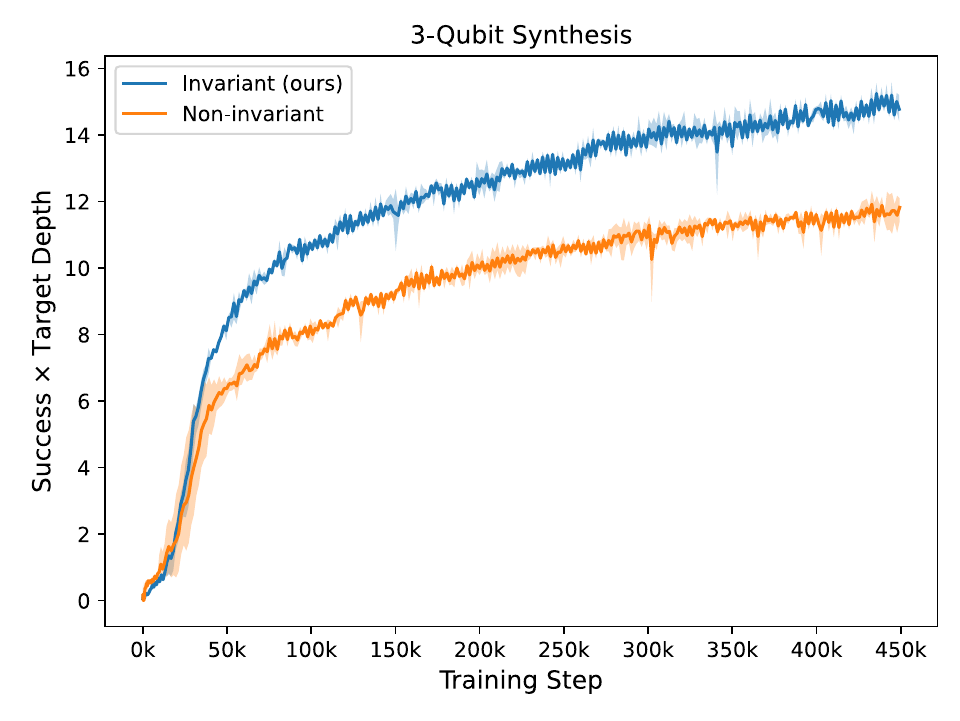}
        \label{fig:ablation_q3}
    \end{subfigure}
    \caption{
        Ablation study for 1-qubit (left) and 3-qubit (right) synthesis.
        Lines are the mean of four runs, shaded regions depict extrema.
        The $y$-axis \emph{scales agent success given a problem instance's difficulty}.
        Global phase invariance can be learned in the simplest cases (1-qubit synthesis), but is necessary for harder problems (2- and 3-qubit synthesis).
        Global-phase invariant architectures outperform non-invariant ones given enough training time.
    }
    \label{fig:ablation}
\end{figure*}

\begin{figure}[t]
    \centering
    \includegraphics[width=0.3\textwidth]{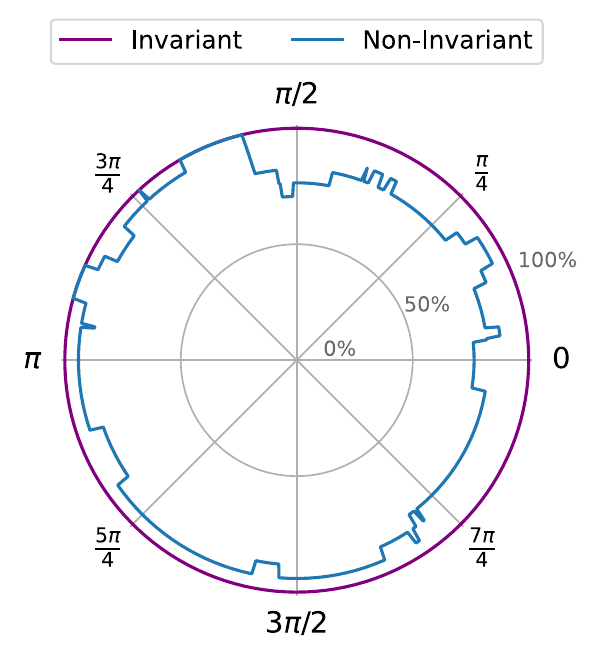}
    \caption{
        Accuracy of global phase invariant and non-invariant architectures as a function of phase shift. Two otherwise identical networks are trained to match unitaries with their associated Clifford+$T$ gates. Global phase shifts (angles on the unit circle) are applied to unitaries during training and testing. The \emph{invariant} architecture maintains perfect accuracy (keeps a radius of $100\%$); it correctly matches unitaries with their associated gate no matter the global phase.
    }
    \label{fig:invariance}
\end{figure}

\section{Experiments}
\label{sec:experiments}

\begin{figure*}[ht]
    \centering
    \includegraphics[width=.75\textwidth]{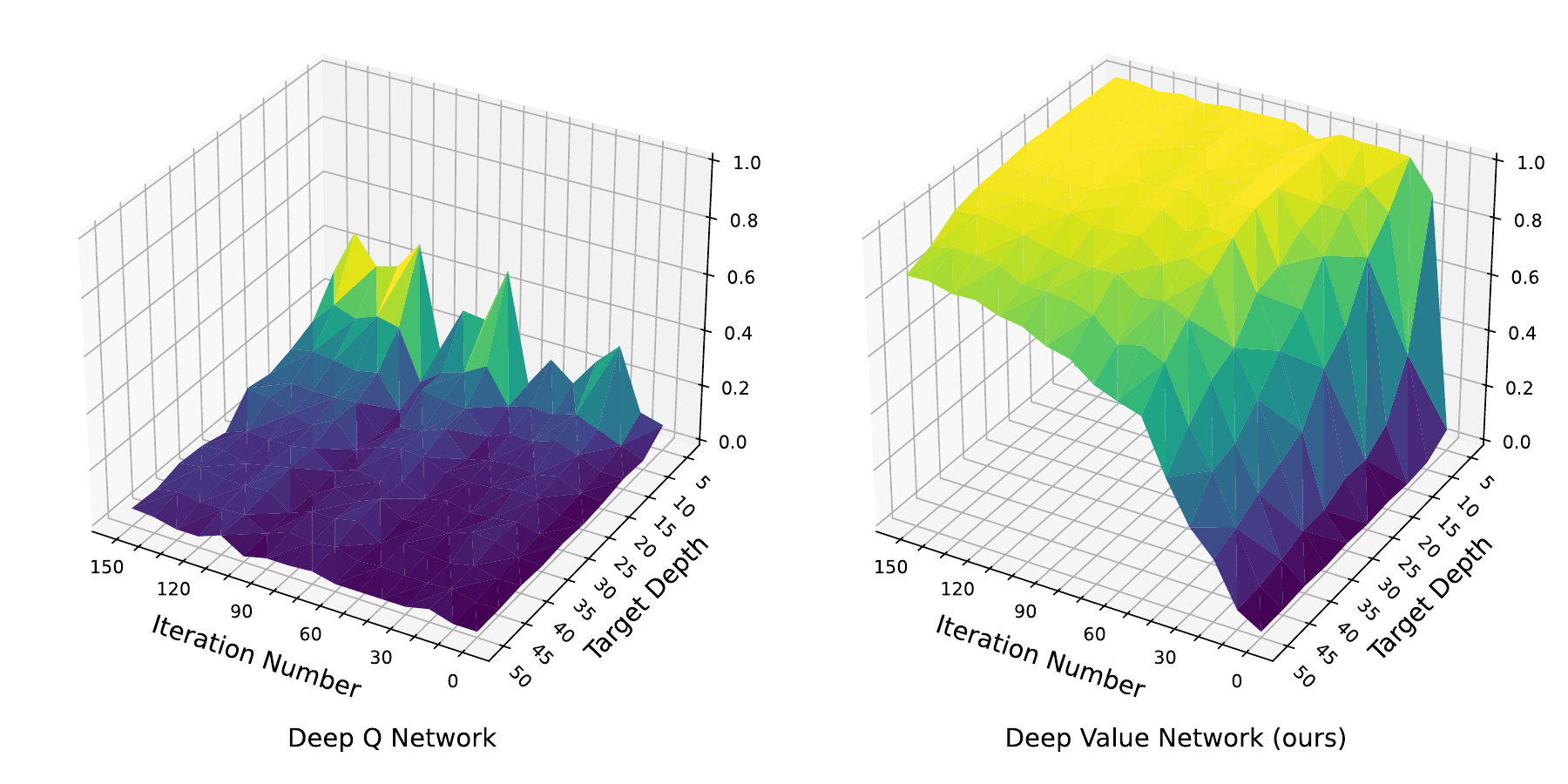}
    \caption{
        Probability that a DQN agent (left) and a DVNU agent (right) successfully finds implementations of random 1-qubit Clifford+$T$ unitaries by inversion. 
        The DQN learns a Q function (Eq~\ref{eqn:qfunc}) whereas the DVNU agent learns a value function (Eq~\ref{eqn:valuefunc}).
        The \emph{Target Depth}-axis shows the number of gates in 250 randomly generated compilation targets. Adding gates increases compilation difficulty.
        The \emph{Iteration Number}-axis measures how much training time has passed.
        The vertical axis shows the probability that a compilation agent will successfully compile a target circuit.
        The architectural modifications made to the DVNU agent improve upon previously demonstrated state-of-the-art deep-learning-based unitary synthesis approaches.
    }
    \label{fig:successcomparison}
\end{figure*}

In this section we evaluate our Reinforcement Learning approach to quantum circuit synthesis detailed in Section~\ref{sec:method}.
We compare various RL synthesis approaches on randomly generated unitaries inputs and a series of test benchmarks.



\subsection{The Impact of the Proposed Architectural Modifications}
How do the proposed neural architectural modifications proposed in Section~\ref{sec:method} impact synthesis performance?
To answer this question, we can compare the ability of global-phase invariant and non-invariant architectures to learn useful properties about unitaries.
We take as a proxy task the problem of predicting which unitary is associated with which gate in a finite (non-parameterized) gate set. 
Two structurally identical multi-layer perceptron \cite{rosenblatt_1958_perceptron} networks are trained to assign gate labels to unitaries associated with the two-qubit Clifford+$T$ gate set for 250 epochs.
Figure~\ref{fig:invariance} illustrates how the network which first applies the global-phase invariance operation detailed in Section~\ref{sec:invariantarch} is resilient to global phase disturbances; no matter the global phase shift it is able to correctly match unitaries with their associated gates.

We emphasize that other than the global-phase invariance operation, the two networks trained in this experiment have identical architectures.
This experiment highlights how global-phase invariance allows for resources which would otherwise be dedicated to learning global-phase invariant representations are free to be used for learning more meaningful qualities of a data set. 
Our experiments indicate that increasing the size of the neural networks while keeping the task the same allows for the performance of the non-invariant architecture to catch up.

This effect can be seen when approaching synthesis problems with different difficulties.
For simpler problems, like single-qubit unitary synthesis, the impact of global-phase invariance is less evident; agents trained with and without dephasing achieve similar performance (as indicated in the left graph in Figure~\ref{fig:ablation}).
For harder, multi-qubit, synthesis problems the need for global-phase invariance is more pronounced (see right graph in Figure~\ref{fig:ablation}).

The impact of the unitary perturbation data augmentation technique and the advantage of using NeRF encodings are both evident even when solving simpler single-qubit synthesis problems (see left graph in Figure~\ref{fig:ablation}).

\subsection{Evaluating DVNU Against Other RL Approaches}
\label{sec:mb_vs_mf}

For this experiment, random unitary targets are obtained by randomly generating quantum circuits in the Clifford+$T$ gate set consisting of $\{H, X, Y, Z, S, T, S^\dagger, T^\dagger \}$ gates.
Circuits are generated by randomly selecting a member of the single-qubit Clifford group followed by either $T$ or $T^\dagger$.
Random circuits are used during training because they can easily be generated and because the difficultly of synthesizing the unitaries associated with these circuits can be tuned by adding or removing gates.
The metric of interest here is the likelihood that a discrete gate synthesis algorithm finds a circuit implementation of a target unitary.
There are $O(|\actionspace|^d)$ depth $d$ possible circuits to check, so performance on this task estimates how well an agent has learned to navigate this huge state space.

We train two models, a DQN (which learns a Q function) as in \cite{moro_2021_drlcompiling, chen_2022_efficient} and a DVNU (which learns a value function).
The DQN is trained without any of the architectural modifications propsed in Section~\ref{sec:method}.
Agents are trained for 150 iterations (about 6 hours) on a machine with an AMD EPYC 7763 CPU and an NVidia A100 GPU. 
Figure~\ref{fig:successcomparison} illustrates the results of this experiment.
After training, the single-qubit DVNU agent was able to synthesize random Clifford+$T$ circuits of depth 80 with a success rate of 54\%.
The DQN with no architectural modifications is only able to synthesize targets containing 15-20 gates with a success rate of 35\%.

Given the same time budget, the DVNU agent achieves higher synthesis performance across all evaluated problem difficulties.
This can partially be explained by the greater sample efficiency of value iteration compared to Q learning.
A deterministic and exact model of synthesis and our proposed architectural modifications enable the DVNU to train far more efficiently.

\subsection{Synthesis-by-Diagonalization with Machine Learning}
\label{sec:quality_comparison}
Recent work has demonstrated how a wider array of unitaries can be successfully synthesized to high precision when the synthesis objective is changed from inversion to diagonalization~\cite{weiden_2025_diagonalization}.
Here, we present results showing how an RL agent using the DVNU architecture can synthesize unitaries taken from real quantum circuits.
Instead of comparing to other RL agents, we compare to the Quantum Shannon Decomposition.
When synthesizing unitaries to precisions $\epsilon \leq 10^{-3}$, pure RL search algorithms fail.

\begin{table*}[ht]
    \centering
    \begin{tabular}{|c|c|c|c|c|c|c|c|c|c||c|}
    \hline
    \multirow{9}{*}{\rotatebox{90}{2 Qubits}} & & Heisenberg & HHL & Hubbard & QAOA & QPE & Shor & TFIM & VQE & Mean \\
    \hline
    & Success Rate & 100\% & 100\% & 100\% & 100\% & 100\% & 100\% & 100\% & 100\% & 100\% \\
    & & 93.7\% & 37.4\% & 42.8\% & 27.1\% & 29.1\% & 41.7\%  & 51.0\% & 20.0\% & 42.9\% \\
    \cline{2-11}
    & T Count  & 495.18 & 530.89 & 443.84 & 579.32 & 551.82 & 511.85 & 430.50 & 341.62 & 485.98\\
    &  & 70.03 & 115.16 & 59.50 & 74.20 & 64.85 & 109.30 & 70.00 & 65.96 & 78.63\\
    \cline{2-11}
    & Clifford Count & 711.66 & 762.20 & 638.63 & 832.15 & 792.35 & 736.77 & 617.06 & 491.33 & 698.27\\
    & & 104.39 & 168.01 & 88.16 & 108.10 & 94.50 & 160.07 & 105.47 & 102.31 & 116.38\\
    \cline{2-11}
    & T Reduction & \bf{85.9\%} & \bf{78.4\%} & \bf{86.6\%} & \bf{87.2\%} & \bf{88.5\%} & \bf{78.7\%} & \bf{83.7\%} & \bf{80.7\%} & \bf{83.5\%} \\
    \cline{2-11}
    \hline
    \hline
    \cline{2-11}
    \multirow{7}{*}{\rotatebox{90}{3 Qubits}} 
    & Success Rate & 100\% & 100\% & 100\% & 100\% & 100\% & 100\% & 100\% & 100\% & 100\% \\
    & & 27.4\% & 9.6\% & 55.6\% & 11.2\% & 78.5\% & 18.0\% & 9.4\% & 13.3\% & 27.9\% \\
    \cline{2-11}
    & T Count  & 3123.46 & 2130.08 & 2572.95 & 2168.42 & 3283.56 & 3058.70 & 2425.20 & 3051.74 & 2726.50\\
    &  & 131.70 & 146.35 & 61.37 & 158.90 & 107.14 & 294.98 & 133.17 & 50.83 & 135.29\\
    \cline{2-11}
    & Clifford Count  & 3601.85 & 2462.84 & 2972.88 & 2506.86 & 3786.61 & 3531.55 & 2800.35 & 3520.56 & 3147.64\\
    &  & 163.46 & 180.74 & 84.45 & 191.82 & 132.40 & 348.72 & 168.76 & 74.29 & 167.78\\
    \cline{2-11}
    & T Reduction & \bf{95.8\%} & \bf{93.2\%} & \bf{97.6\%} & \bf{92.7\%} & \bf{96.8\%} & \bf{90.4\%} & \bf{94.5\%} & \bf{98.3\%} & \bf{95.1\%} \\
    \cline{2-11}
    \hline
    \end{tabular}
    \caption{
        Synthesis of 2- and 3-qubit unitaries taken from partitioned circuits (see Figure~\ref{fig:transpilation} for an illustration).
        In both cases, each unitary is synthesized in approximately 1 second.
        When $R_Z$ gates appear due to the diagonalization process, they are synthesized using \emph{gridsynth}~\cite{ross_2016_gridsynth}. Each $R_Z$ gate is synthesized to a precision of $\epsilon=10^{-7}$ which results in approximately 70 T gates and 110 Clifford gates per $R_Z$.
        This table reports the success rate, Clifford, and non-Clifford T gate counts, and the T count reduction compared to the Quantum Shannon Decomposition (QSD) for unitaries taken from a variety of benchmarks.
        Compared to the QSD (top rows), the diagonalizing agent (bottom rows) on average reduces the number of $R_Z$ gates by 83.5\% and 95.1\% for 2- and 3-qubit unitaries respectively. 
        Comparisons are made to the QSD because other synthesis tools fail to find solutions when $\epsilon \leq 10^{-3}$.
    }
    \label{tab:qsd_comparison}
\end{table*}
Given a 2- or 3-qubit target unitary, our goal is to evaluate how effectively the DVNU-based diagonalization approach can synthesize it.
For all experiments, we require synthesized circuits to approximate the target within a distance of $\epsilon = 10^{-6}$ (see Equation~\ref{eqn:distance}).
In practice, the diagonalizing agent often produces implementations with far higher precision, and this particular choice of $\epsilon$ is largely arbitrary.

As a reference, we compare the DVNU diagonalizer to the Quantum Shannon Decomposition (QSD) \cite{shende_2006_qsd}, an analytical method selected because other search-based tools failed to reach comparable precision.
QSD results are optimized by merging $R_Z$ gates before converting into the Clifford+$T$ gate set when possible.

To train the DVNU diagonalizer, we randomly generated target unitaries of the form $A R_Z^{\otimes n}(\theta_i) B$, where $A$ and $B$ are independently sampled Clifford+$T$ circuits, and $\theta_i$ are randomly selected angles.
For 2-qubit targets, $A$ and $B$ combined to at most 30 gates; for 3-qubit targets, up to 20 gates.

\begin{figure}[h]
    \centering
    \includegraphics[width=\linewidth]{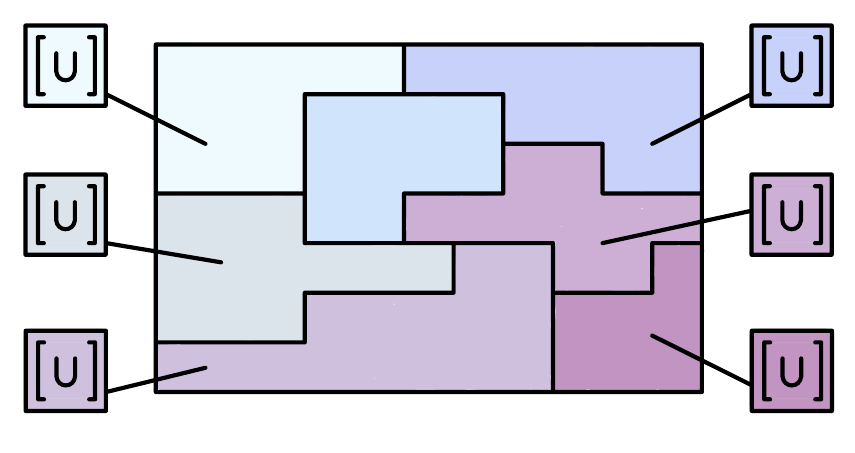}
    \caption{
        A quantum circuit partitioned into many subcircuits.
        The unitaries associated with each one of these subcircuits can be considered independently.
        The results presented in Table~\ref{tab:qsd_comparison} evaluate the DVNU based diagonalizing agent using a circuit partitioning method like this.
        The partitioning algorithm used is the \emph{Quick Partitioner} implemented in the BQSKit toolkit~\cite{younis_2021_bqskit}.
    }
    \label{fig:transpilation}
\end{figure}

We evaluated performance on unitaries extracted from partitioned quantum algorithms (see Figure~\ref{fig:transpilation} for the partitioning process).
This suite includes Shor’s algorithm \cite{shor_1997_shorsalgorithm}, TFIM, Heisenberg, and Hubbard model circuits from quantum chemistry simulations \cite{bassman_arqtic_2021}, as well as trained VQE \cite{peruzzo_2014_vqe}, QAOA \cite{farhi_2014_qaoa}, and Quantum Phase Estimation circuits \cite{kitaev_1995_qpe}.
VQE and QAOA benchmarks were generated using MQTBench \cite{quetschlich_2023_mqtbench}.
All unitaries were filtered to ensure uniqueness, yielding 22,323 2-qubit and 45,202 3-qubit instances.

Table~\ref{tab:qsd_comparison} summarizes the results. The DVNU agent successfully synthesized 42.9\% of 2-qubit unitaries and 27.9\% of 3-qubit unitaries across all benchmarks.
While success rates decrease with circuit complexity, the synthesized circuits consistently require far fewer non-Clifford gates compared to QSD.
On average, the DVNU approach reduces the number of non-Clifford gates by 83.5\% for 2-qubit unitaries and 95.1\% for 3-qubit unitaries.

Compared to the QSD, the diagonalizing agent produces solutions with far fewer non-trivial rotation gates.
Compared to the QSD, the DVNU synthesizer reduces the average number of non-Clifford gates by $83.5\%$ for two qubit unitaries and $95.1\%$ for three qubit unitaries.
Although analytical decompositions achieve higher success rates, the substantial reduction in non-Clifford gate count and the efficiency of the DVNU synthesis process highlight its practical advantages.

\section{Discussion}
\label{sec:discussion}
In this section, we offer interpretations and insight into the values which are learned by the Deep Value Network for Unitaries.
Although Reinforcement Learning based synthesis agents are incapable of synthesizing all unitaries, we argue for their utility in scenarios where fast inference is necessary and some amount of failure is tolerable.

\begin{figure*}[]
    \centering
    \begin{subfigure}[t]{\columnwidth}
        \centering
        \includegraphics[width=0.8\columnwidth]{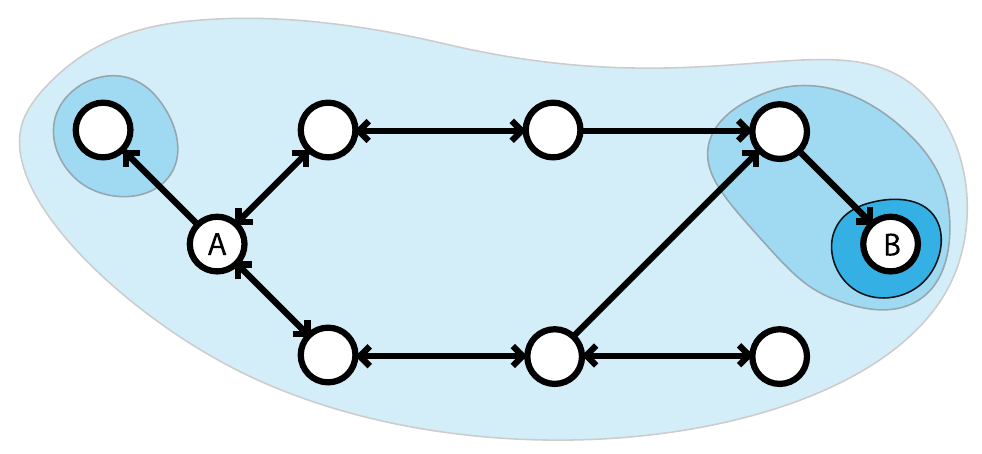}
        \caption{State transition diagram under a noisy and flat value function.}
    \end{subfigure}
    \begin{subfigure}[t]{\columnwidth}
        \centering
        \includegraphics[width=0.8\columnwidth]{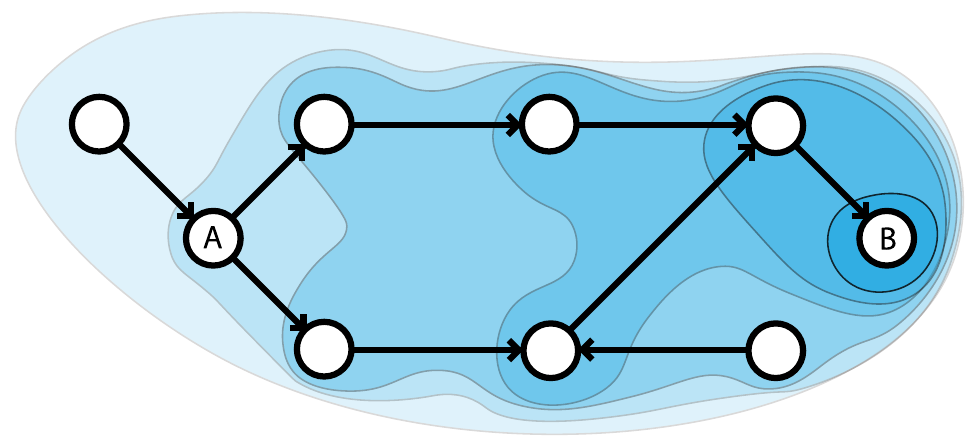}
        \caption{State transition diagram under the learned value function.}
    \end{subfigure}
    \begin{subfigure}[t]{0.48\textwidth}
        \centering
        \includegraphics[width=0.9\textwidth]{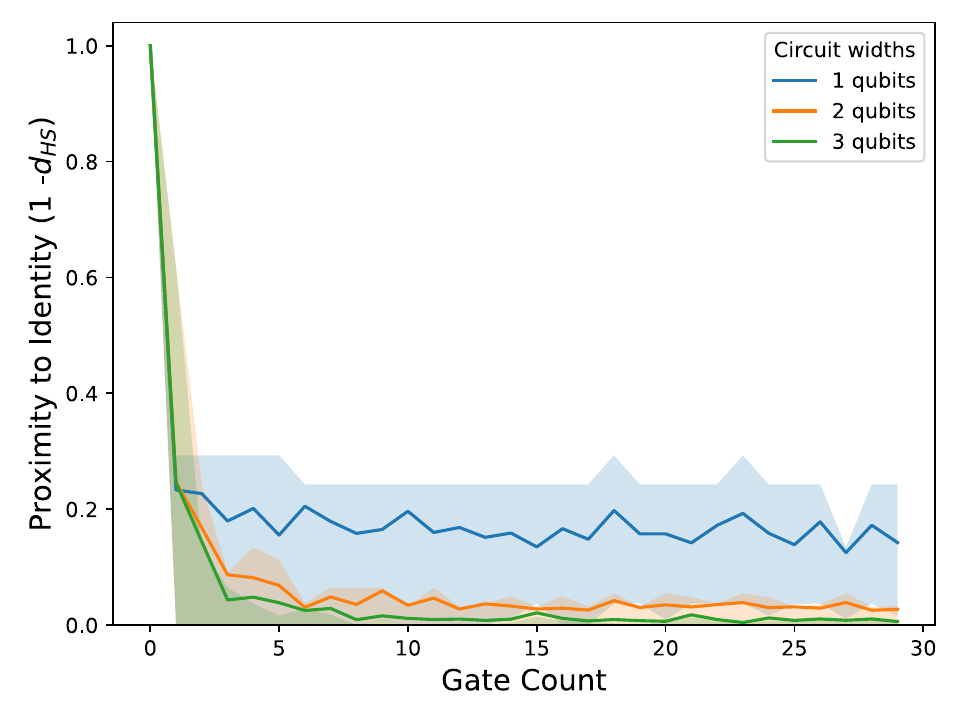}
        \caption{
            Proximity to the identity ($1 - d_{HS}(s_t, I)$) vs. gate count of random Clifford+$T$ circuits.
            This value function is noisy and flat, causing agents to act randomly in most of the state space.
        }
    \label{fig:depthvdistance}
    \end{subfigure}
    \hspace{4pt}
    \begin{subfigure}[t]{0.48\textwidth}
        \centering
        \includegraphics[width=0.9\textwidth]{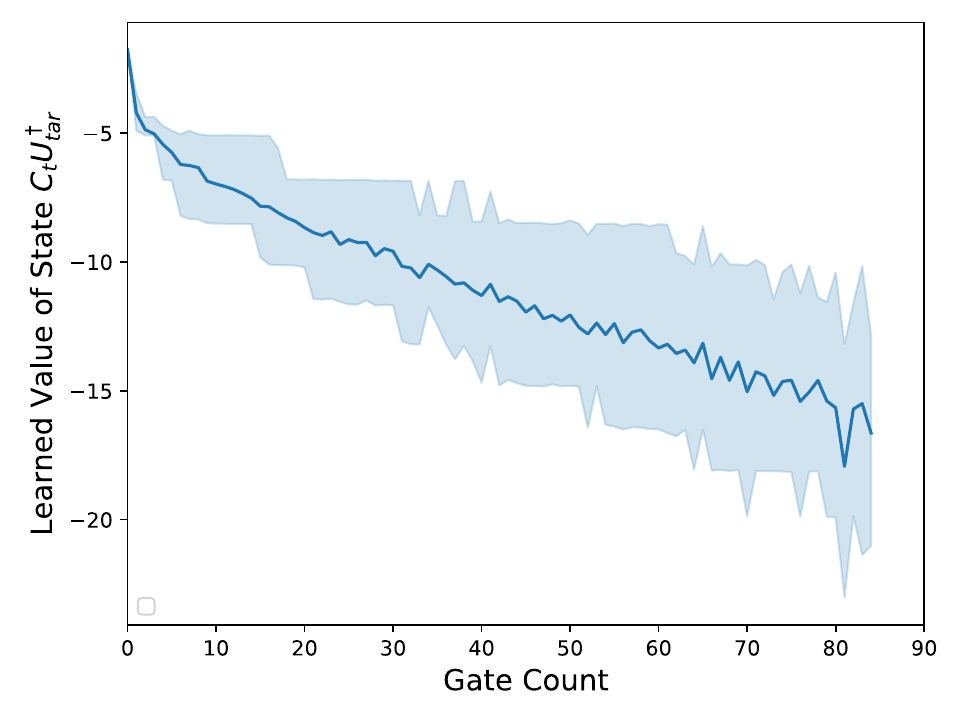}
        \caption{
            Learned value of synthesis states vs. gate count of random single qubit Clifford+$T$ circuits.
            This value function guides agents towards terminal states.
        }
        \label{fig:depthvvalue}
    \end{subfigure}
    \caption{
        The impact of the value function on state transitions.
        (a) Agents which use noisy and flat value functions to plan actions will act randomly.
        Nodes $A$ and $B$ represent the initial and terminal states respectively.
        Darker contours represent increasing value magnitudes. 
        (b) Agents which use learned values to plan actions tend towards terminal states.
        (c) shows the Hilbert-Schmidt distance which is a noisy and flat value function.
        Lines are the mean distances of 250 random circuits, the shaded regions are interquartile ranges.
        In synthesis with discrete gate sets, unitary distance does not correlates with gate count.
        States which require fewer gates to reach a terminal state generally do not correspond to lower distances than states which require more gates. 
        (d) shows the learned value function which is based on path distance.
        States which require fewer gates to reach a terminal state have higher values than those which require more gates.
        Gate cancellation in randomly generated circuits contributes to variance in state values.
    }
\end{figure*}
\subsection{Interpreting the Value Function}
In order for a function to meaningfully guide search-based synthesis,
it should correlate with the goal of finding short gate count
implementations of unitaries.  Intuitively, the value of a synthesis
state should depend on the number of gates needed to transform that
state into a terminal state.

A discrete gate set synthesis algorithm which relies directly on the
Hilbert-Schmidt distance function (or Frobenius norm, diamond norm, etc.) to guide the placement of gates will
randomly search the space of possible circuits.
Many synthesis algorithms to date use these cost functions to guide gate placement \cite{davis_2020_qsearch, smith_2021_leap, paradis_2024_synthetiq}.
Figure~\ref{fig:depthvdistance} shows how for random Clifford+$T$
circuits, Hilbert-Schmidt distance does not correlate strongly with
the number of gates in the circuit (and therefore the number of gates
needed to synthesize that unitary).  Synthesis states in the
neighborhood of the current state appear equally distant from a
terminal state.

\emph{How does a Reinforcement Learning agent avoid this pitfall?} Under a reward function such as Equation~\ref{eqn:reward}, the learned value of states correlates with the number of gates in the circuit.
As shown in Figure~\ref{fig:depthvvalue}, the value of a state is a \emph{learned, scaled and, shifted} representation of the \emph{gates to go}.
This property makes the agent a fast and efficient synthesizer; it ideally takes the shortest direct path to a terminal state.

In the case of exact synthesis of single-qubit Clifford+$T$ unitaries, the \emph{smallest divisor exponent} (SDE) plays a similar role as the value function \cite{kliuchnikov_2012_exact}.
This synthesis algorithm sequentially takes actions depending on which gate decreases the SDE.
In the multi-qubit case, it is believed that such a simple function does not naturally exist \cite{gosset_2013_tcount}.
The model-based RL approach taken in this paper can be construed as a learned version of the SDE function for approximate multi-qubit synthesis. 

\subsection{Suitability for Other Tasks}
The architectural modifications proposed here are designed specifically for the task of synthesizing quantum circuits using discrete gate sets.
Although we consider only the Clifford+$T$ gate set here, in principle, any discrete universal gate set can be used (e.g. Clifford+V, Clifford+$\sqrt{T}$, or others).
In our tests, the DVNU's architectural modifications presented in this paper benefit ML-based synthesis for these alternative gate sets as well.

Another task of interest to the quantum compiler community is that of \emph{ansatz classification}: pairing a unitary matrix with a parameterized circuit ansatz which is likely to be capable of implementing it.
Several recent works have used ML to solve this problem~\cite{weiden_2023_qseed, kremer2024aimethodsapproximatecompiling}.
In our tests, the modifications presented in this paper offered no benefit compared to an architecture not using our modifications.
We trained two models (one with modifications, one without) to determine which of 20 possible parameterized ansatzes could have generated 100,000 different random training samples.
Unitary perturbations allows for regions around a unitary exactly implementable in the Clifford+$T$ gate set to be associated with values.
This blurring results in an effect like as shown in Figure~\ref{fig:invar_depiction}.
However, the process of randomly sampling unitaries that can be implemented by a parameterized ansatz (i.e. randomly setting ansatz parameters) already shows the network representative points from the whole region of implementable unitaries.

Ansatz classification and synthesis are also differentiated by how values are learned.
In synthesis, an agent is trained with reinforcement learning to recognize \emph{good} states.
The training process here requires that the network know that low depth circuits are \emph{good} before longer depth circuits can be accurately considered.
This bootstrapped style of learning can be instable~\cite{suttonbarto}.
In ansatz classification, a synthetic dataset of unitaries can be generated, and \emph{supervised learning} can be used.
Supervised learning does not have the same boostrap instability as reinforcement learning.
The comparative ease and stability of the supervised learning task of ansatz classification helps explain why global-phase invariance was less helpful. 
The results highlighted in Figure~\ref{fig:ablation} support this finding, the easier task of learning to synthesize single-qubit unitaries did not benefit from using global-phase invariance.
The harder task of three-qubit synthesis did however.

\subsection{Monte Carlo Tree Search Synthesis}
Monte Carlo Tree Search (MCTS) is a popular approach to solving problems which require searching through a large discrete state space~\cite{suttonbarto}.
Recent work has applied the MCTS to the unitary synthesis problem~\cite{rietsch_2024_unitary}.
As common implementations of MCTS require agents to learn both policies and value functions, MCTS synthesis agents are likely to see benefits from using our architecture.

\section{Conclusion}
\label{sec:conclusion}
In this paper, we have introduced modifications to basicneural networks: global phase invariance and unitary perturbations.
These modifications enable deep learning techniques to more effectively synthesize circuits which approximately implement unitary matrices.
Along with an encoding technique adopted from 3D image processing, we propose a network architecture called the Deep Value Network for Unitaries (DVNU).
The effectiveness of our Reinforcement-Learning-based approach is demonstrated using ablation studies (i.e. selectively removing modifications then training as usual). 
We also compare to Deep Q Networks which have been used for this task several times in the literature. 
As we have only proposed architectural modifications, other RL approaches for quantum circuit synthesis, such as Monte Carlo Tree Search can adopt these ideas as well.
We believe that our work will inspire more research into neural architectures specialized for tasks of interest in the quantum compilation field.



\bibliographystyle{IEEEtranS}
\bibliography{refs}

\end{document}